\documentclass[12pt,a4paper]{lijuan}

\title{Dominance and multi-locus interaction}
\author[1,2,a]{Juan Li}
\author[1,2]{Claudia Bank}
\affil[1]{Institute of Ecology and Evolution, University of Bern, Bern, Switzerland}
\affil[2]{Swiss Institute for Bioinformatics, Lausanne, Switzerland}
\affil[a]{To whom correspondence should be addressed: lijuan.a@outlook.com}

\date{ } 

\begin{document}

\maketitle

\section*{Abstract}

Dominance is usually considered a constant value that describes the relative difference in fitness or phenotype between heterozygotes and the average of homozygotes at a focal polymorphic locus. However, the observed dominance can vary with the genetic background of the focal locus. Here, alleles at other loci modify the observed phenotype through position effects or dominance modifiers that are sometimes associated with pathogen resistance, lineage, sex, or mating type. Theoretical models have illustrated how variable dominance appears in the context of multi-locus interaction (epistasis). Here, we review empirical evidence for variable dominance and how the observed patterns may be captured by proposed epistatic models. We highlight how integrating epistasis and dominance is crucial for comprehensively understanding adaptation and speciation.

Keywords: Dominance modifier, epistasis, fitness landscape, polygenic traits, heterozygote, hybrid incompatibility.

\section{The complex interplay of dominance and epistasis}

Genetic polymorphism is ubiquitous and contributes to the diversity of phenotypes and fitness in natural populations. A polymorphic locus can contain multiple alleles (see \hyperlink{glossary}{Glossary}) with different effects on phenotype and fitness. When two different alleles appear in a heterozygous genotype, their combined effect may deviate from the expected average effect, resulting in a phenomenon known as dominance. The evolution and mechanisms of dominance have been debated in genetics and evolutionary biology for more than a century (summarized in Box \hyperlink{history}{1}; reviewed in \cite{mayo1997evolution, bourguet1999evolution, bagheri2006unresolved}). In classical Mendelian genetics, dominance describes the masking effect between two alleles at a single locus, where one allele (the dominant allele) masks the effect of the other allele (the recessive allele). The masked allele may exhibit incomplete (partial) dominance or recessivity. However, it has long been proposed that dominance is determined by genetic interaction between loci (also termed epistasis; see \hyperlink{glossary}{Glossary}) rather than the sole interaction of two alleles at a focal locus \cite{wright1929fishera}.

In the realm of genetic interactions, dominance and epistasis often intersect and interfere with each other \cite{billiard2021integrative}. Indeed, the concept of epistasis was first introduced to explain variable dominance, namely when the dominance at one locus depends on the genetic background \cite{bateson1910reports}. Here, the term genetic background referred to a second polymorphic locus in a Mendelian system. Nowadays, the term epistasis engulfs any non-genetic interaction between at least two loci that non-additively affects phenotype or fitness and is often considered distinct from dominance. The relationship between dominance and epistasis has sparked ongoing debates and challenges in defining and conceptualizing them within a unified framework. 

Here, we first summarize the classical genetic viewpoints of dominance, specifically, constant dominance and position effects. Second, we review examples that illustrate variable dominance as a consequence of epistasis. Thirdly, we discuss how constant and variable dominance can be accounted for in one model, Wade's three-locus model. Finally, in the scope of how dominance contributes to multi-locus interactions and vice versa, we discuss the potential for incorporating dominance and epistasis into fitness landscape models (see \hyperlink{glossary}{Glossary}). 
\FloatBarrier

\begin{glossary}[ht]
~
\hypertarget{glossary}{\caption{{\textbf{Glossary}}}}
\begin{center}
\begin{minipage}{\textwidth}
\flushleft
\small

\textbf{Additive effect} is the contribution of an allele to phenotype or fitness that is independent of other genetic factors.

\textbf{Dominance} describes the relative difference in fitness or phenotype between a heterozygote and the average of the two corresponding homozygotes at a focal locus.

\textbf{Dominance coefficient ($h$)} is a parameter that quantifies the level of dominance at a locus. This coefficient $h$ can be zero, positive or negative. $h=0.5$ means no dominance, i.e., when two alleles are additive. $h=0$ describes recessivity, i.e., when the effect of a focal allele is completely masked by the other allele(s). $h=1$ when the focal allele is completely dominant. $h\neq 0.5$ and between 0 and 1  is termed incomplete (partial) dominance. $h<0$ is termed under-dominance; here, the trait/fitness value of the heterozygote is lower than the values of both homozygotes. $h>1$ is termed over-dominance; here, the trait/fitness value of heterozygotes is higher than the values of both homozygotes.

\textbf{Dominance modifier} is an allele the presence of which changes the dominance at another locus.  

\textbf{Epistasis} is the interaction among alleles at two or more loci resulting in non-additive effects on fitness or phenotype.

\textbf{Fitness landscape} is a map of genotypes to fitness. 

\textbf{Locus} refers to a continuous region in a genome that encodes a specific function. A locus can be a nucleotide site, an amino-acid site, a gene, or a regulatory element. Genetic variants at the same locus are referred to as \textbf{alleles}.

\textbf{Non-additive effect} is the specific contribution of a group of alleles at the same or different loci to phenotype or fitness. There are two types of non-additive effects, epistasis (non-additive contribution of alleles at different loci) and dominance (non-additive contribution of alleles at the same locus). 

\textbf{Polygenic trait} is a trait, the value of which is determined by many loci. 

\end{minipage}
\end{center}
\end{glossary}

\begin{supText}[ht]
~
\hypertarget{history}{\caption{{\textbf{Box 1 -- Evolution and mechanisms of dominance}}}}
\begin{center}
\begin{minipage}{\textwidth}
\flushleft
\small
\textbf{Fisher's hypothesis} ~~ Fisher proposed that new alleles are initially co-dominant and that the evolution of dominance occurs through the step-wise accumulation of dominance modifier mutations (\hyperlink{glossary}{Glossary}). According to Fisher, dominance modifiers gradually shift the dominance level of an allele, making it either more dominant or more recessive \cite{fisher1928possible}. He assumed that the modification effects were weak. Fisher's hypothesis is debated because the advantage of a modifier mutation depends on the presence of the allele that is modified, resulting in (even weaker) second-order selection. Moreover, the emergence of dominance modifiers is constrained by low mutation rates \cite{wright1929fishera, wright1929fisherb}. These two factors imply a long waiting time for a dominance modifier to fix in the population. Because of the long waiting time in combination with the proposed small effects of modifiers, Fisher's model is not well-suited to explain why many loci exhibit large deviations from additivity. \cite{wright1929fishera,wright1929fisherb}.

\textbf{Wright's physiological explanation} ~~  Wright promoted a physiological explanation for dominance based on the catalytic process of enzymes \cite{wright1934physiological}. According to Wright, in the catalytic process, the value of a trait depends on the combined effect of a series of loci. Each locus is only in part responsible for the trait. The relationship between the effect of a heterozygous genotype and the trait is indirect. Therefore, the observed dominance at a focal locus varies depending on the variation at other loci that contribute to the trait. Wright's explanation was expanded in subsequent studies of genes that belong to the same pathway, e.g., \cite{kacser1981molecular,kondrashov2004common}.

\textbf{Empirical evidence points to widespread recessivity of new mutations} ~~ The observed dominance of mutations appears to be correlated with their fitness consequences. New mutations tend to be deleterious and recessive. Across multiple organisms, studies have consistently estimated the dominance coefficients of new mutations to be around $0.2-0.25$ on average \cite{mukai1972mutation, agrawal2011inferences, manna2011fitness, yang2017incomplete}. As the deleterious effect of mutations increases, they tend to be more recessive \cite{charlesworth1979evidence, phadnis2005widespread, huber2018gene}. Indeed, new mutations seem to display suboptimal expression relative to their ancestral variants, whose expression may have adjusted to the given genetic background \cite{huber2018gene}. Thus, consistent with Wright's theory, the observed recessivity of new mutations seems to be a by-product of the inherent nature of gene regulatory networks that renders new mutations suboptimal rather than a consequence of prolonged evolution. For instance, in a species of algae (\textit{Chlamydomonas reinhardtii}), which predominantly consists of haploid individuals that can form temporary diploid individuals, mutations are recessive just as often as among permanently diploid species, \textit{e.g.}, \textit{Drosophila} \cite{orr1991test}.

\end{minipage}
\end{center}
\end{supText}

\begin{figure}
    \centering
    \includegraphics[width=0.92\textwidth]{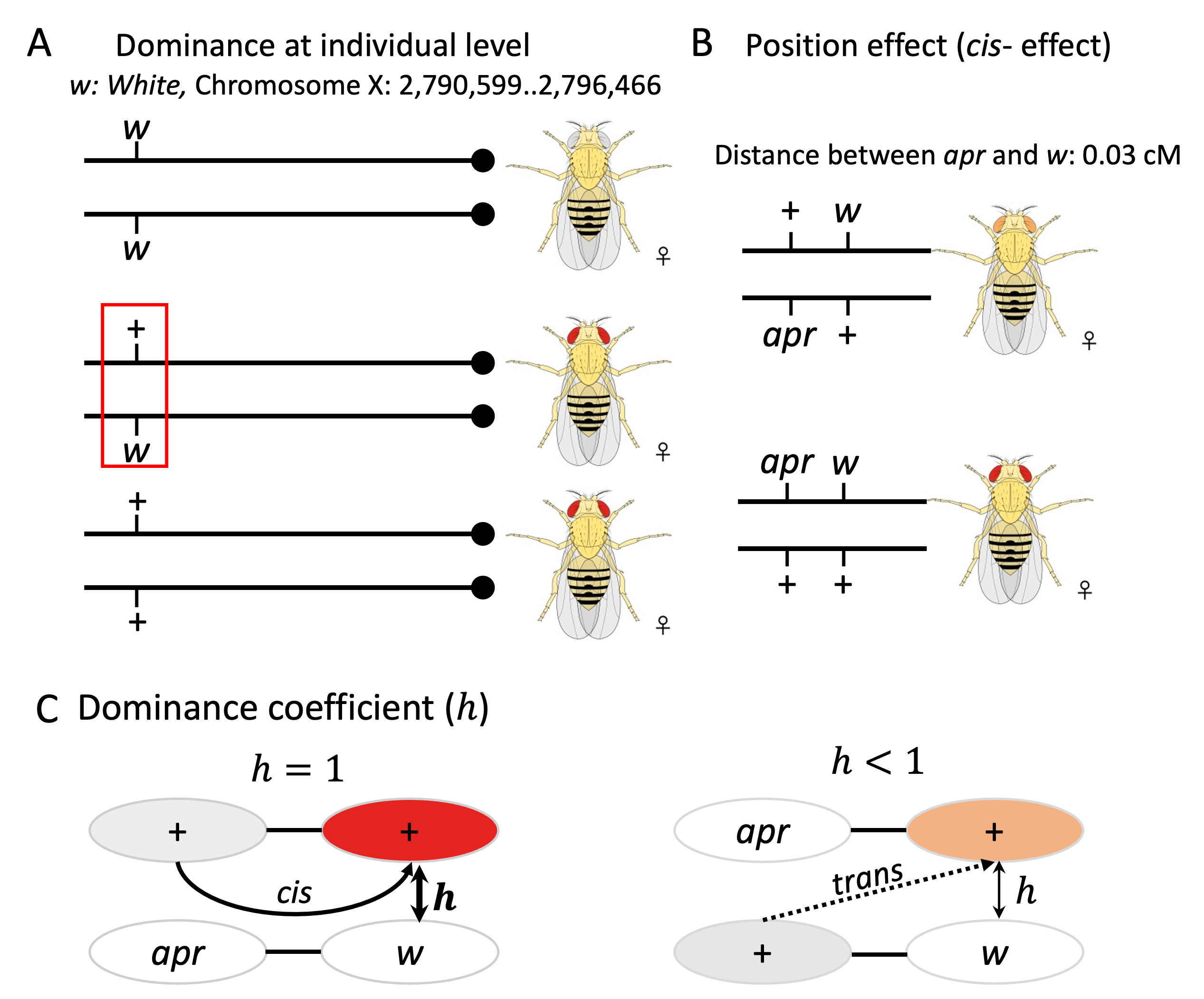}
    \caption{Classical definition of dominance. A. Constant dominance at one locus with Mendelian segregation. The white-red eye system in \textit{Drosophila melanogaster} was mapped to one gene, \textit{white} \cite{morgan1911origin}. The wild-type allele ($+$) of gene \textit{white} is dominant over its mutant ($w$) regarding the expression of red eye pigmentation. B. Position effect or variable dominance due to polymorphic \textit{cis}-regulatory elements. A \textit{cis}-regulator, \textbf{apr}, affects the red eye color in heterozygotes \cite{lewis1952pseudoallelism}. The red eye color only presents when the wild-type allele ($+$) at locus \textbf{apr} is linked to the wild-type allele at \textbf{white} locus ($apr~w / +~+$; $apr~w$ on one chromosome; $~+~+$ on the other chromosome; the slash $/$ separates the two haplotypes); otherwise, the eye color becomes apricot or yellowish pink ($+~w / apr~+$). C. Schematic illustration of dominance in the case of the position effect in panel B. The wild-type allele $+$ at the \textbf{apr} locus enhances the effect of the wild-type allele $+$ at the \textit{white} locus in \textit{cis}, leading to complete dominance ($h=1$) (left side). Dominance at the \textit{white} locus becomes incomplete ($h<1$) when the wild-type alleles at \textit{white} and \textit{apr} are not on the same chromosome (right panel).}
    \label{domiClassical}
\end{figure}

\begin{supText}[ht]
~
\hypertarget{polygenic}{\caption{{\textbf{Box 2 -- The challenge of inferring dominance at loci involved in polygenic traits}}}}
\begin{center}
\begin{minipage}{\textwidth}
\flushleft
\small
Many traits are determined by the combined effects of alleles at multiple loci. Unlike single-gene traits, in which a genotype at a focal locus in a diverse population is expected to display the same trait (Figure \ref{domiClassical}A; note that we ignore the potential effects of measurement error and phenotypic plasticity for simplification purposes), such polygenic traits (see \hyperlink{glossary}{Glossary}) may exhibit trait variation due to additive and non-additive effects at other segregating loci in the genome. In the context of polygenic traits, most empirical research has focused on investigating dominance at individual loci. The dominance coefficient ($h$, (see \hyperlink{glossary}{Glossary})) is here inferred from the phenotypic variance due to dominance and additive genetic variance observed at a focal locus \cite{comstock1948components, robinson1949estimates}. 

When estimating the effect of additivity and dominance at a segregating locus, it is ideal for their variances to be independent. However, in practice, dominance and additive genetic variance interact and influence each other. Most data and theories support that mainly additive genetic variance contributes to complex trait evolution, but also strong non-additive effects have been observed (\textit{e.g.}, \cite{hill2008data}). A key challenge lies in accurately accounting for dominance in the estimation process, because current methods may create a bias towards allocating non-additive variance to additive variance \cite{crow2009introduction, huang2016genetic}. When the additive effect from two alleles is calculated by assuming no dominance ($h=0.5$), the dominance variance is deprecated. Conversely, when one allele is assumed to be completely dominant ($h=1$), the dominance variance captures the majority of total genetic variation. Thus, depending on the null model, the inferred relative contributions of additive and dominance variances to the overall genetic variation will vary.


The variance analysis described above assumes a normal distribution of trait values for each genotype at the focal locus in a population \cite{huang2016genetic}. Epistasis between segregating alleles can cause a deviation from the expected normal distribution, potentially obstructing inference. Finally, the dominance of epistasis, as discussed in Section \ref{defineDMI}, may further influence observed trait distributions and the possibility of accurate inference.

\end{minipage}
\end{center}
\end{supText}

\section{Constant dominance: the relative effects of alleles at one locus}

\subsection{Dominance in Mendelian genetics}
\label{constantOneLocus}

\begin{displayquote}
IN cultures of \textit{Drosophila ampelophila} that had been closely inbred for a year, a mainly fly, lacking the red pigment of the eye, appeared. The same stock has continued to produce those white-eyed mutants always of the male sex. A white-eyed father transmits the character to about one fourth of his grandsons, but none of his granddaughters. In this sense the character is sex limited. The white eye can be transmitted, however, to the females, most readily by breeding any white-eyed males male to red hybrids (F\textsubscript{1}) out of white by red. White-eyed males and females give pure stock. When a white-eyed female is bred to any wild male all of the female offspring have red eyes and all of the male offspring white eyes. 

\cite{morgan1911origin}
\end{displayquote}

The above quote illustrates Mendelian segregation of a sex-linked gene. In this example, the mutant allele results in a loss of function with a recessive inheritance pattern in the population (Figure \ref{domiClassical}A). Here, \textit{Drosophila ampelophila} is the synonym for the widely studied model organism \textit{Drosophila melanogaster}, which has greatly contributed to the study of genetics. The gene responsible for red eyes, known as white (\textbf{w}), was named based on its mutant phenotype long before its sequence information was available. Now we know that \textbf{w} is located on chromosome X:2,790,599..2,796,466 in the reference genome release 6. This example highlights a (seemingly) perfect map between one gene and a trait. 

\subsection{Position effect (\textit{cis}- effect)}

The position effect refers to \textit{cis}-regulation in short range on a chromosome. The variable dominance caused by the position effect was first investigated at the white (\textbf{w}) locus in \textit{D. melanogaster}. The red-eye allele ($w$) at the \textbf{w} locus was initially reported to be completely dominant (see \ref{constantOneLocus}) \cite{morgan1911origin}. However, subsequent experiments revealed the presence of variable red eye colors (apricot and red) in heterozygotes at the \textbf{w} locus ($w~/~+$). This variation in eye color was subsequently attributed to the regulatory effect from another polymorphic locus, \textbf{apr}. Considering the nearby regulatory locus, the white-red eye pigmentation system was expanded to the white-apricot-red system (Figure \ref{domiClassical}B; \cite{lewis1952pseudoallelism}). The \textbf{apr} locus is closely linked to the \textbf{w} locus at a distance of 0.03 centimorgan (cM) and regulates \textbf{w} in \textit{cis}, i.e., the location of the two alleles on the same DNA molecule matters for the phenotype. In female double heterozygotes, different eye colors are observed: females with the genotype $apr~w~/~+~+$ ($apr~w$ on the one chromosome; $~+~+$ on the other chromosome) display the wild-type red eye color, whereas females with $apr~+~/~+~w$ ($apr~+$ on the one chromosome; $~+~w$ on the other chromosome)  display an apricot eye color. In the latter case, the mutant \textit{apr} does not efficiently regulate the expression of red eye pigmentation, resulting in the apricot eye color. This system illustrates how a modifier allele can change dominance through short-range interaction.  The regulation at the \textbf{w} locus was studied intensively through the disruption of its regulatory regions using mutagenesis or transposable elements \cite{lindsley2012genome}. These investigations have discovered additional dominance scenarios, such as a different mutant allele that is dominant over the wild-type allele \cite{bingham1980regulation}. Regulatory interactions are likely an important contributor to variable dominance. However, they are challenging to detect phenotypically, particularly when the allele-specific regulation effects of a \textit{cis}-element are less striking than in the example above \cite{porter2017new}.

\FloatBarrier
\section{Variable dominance, dominance modifiers and multi-locus interactions} \label{domiEpi}

As shown in the above example, the observed dominance at a locus can vary due to epistatic interactions with alleles at other loci. Such dominance modifiers may be alleles at closely linked regulatory loci, or they may be interacting alleles elsewhere in the genome. If the focal locus is only modified by an allele located on the same DNA molecule, the modification effect is in \textit{cis}, resulting in the position effect. If the effect of the dominance modifier is independent of its physical location, the effect is in \textit{trans}. However, the dominance modification at a focal locus is not limited to a single modifier in either \textit{cis} or \textit{trans}. When one or more modifier alleles are segregating in the population, they can lead to a variable observed dominance coefficient at a focal locus (see empirical examples in Table \ref{varDomi}). Because experimental studies are commonly carried out in individuals with multiple polymorphic loci, the modifying effects on the focal locus are often not discernible as purely \textit{cis} or \textit{trans} interactions, and the dominance-modifying loci are difficult to infer. Instead, the dominance at the focal locus is often inferred as an average over all genetic backgrounds in the population (see Box \hyperlink{polygenic}{2}).  

One notable example of dominance modifiers can be traced back to Sewall Wright's studies of the coat color of guinea pigs\cite{wright1927effects} (Table \ref{varDomi}). He classified the genetic basis of color pigmentation based on different categories, one of which is known as the albino series. In this series, the causal locus determining the grade of brown color consists of a dominant allele $C$ and several recessive $c$ alleles. However, the expression of this locus is regulated by multiple other loci, including a modifier with two alleles ($P$ \& $p$). Only when allele $C$ and its modifier allele $P$ were present, the darkest brown fur was observed. In contrast, individuals with recessive $c$ alleles or modifier genotype $pp$ displayed variable lower shades of brown fur. 

The guinea pigs' pigmentation phenotype is a polygenic trait determined by a large number of genotypes, where alleles at multiple loci interact non-additively. As far as Sewall Wright's extensive exploration of this system went \cite{wright1927effects}, ``the possible combinations among these factors account for 43,740 genetically distinct sorts of guinea pigs. The effects of any given factor can thus be studied on a great variety of genetic backgrounds."  This wealth of genetic variation and the observed importance of gene interactions inspired Sewall Wright to propose one of his most influential concepts, the fitness landscape \cite{wright1932roles}. However, it remains a challenge to simultaneously integrate dominance and epistasis in a single fitness landscape model. To highlight the need for such models, the next section features empirical examples demonstrating emerging evidence of variable dominance caused by multi-locus interactions (Table \ref{varDomi}).

\begin{table}
    \caption{Examples of variable dominance}
    \begin{threeparttable}
    
    \small
    \begin{tabular}{p{0.2\textwidth} p{0.16\textwidth} p{0.12\textwidth} p{0.12\textwidth} p{0.11\textwidth} p{0.12\textwidth}}
      \hline 
      & Causal allele →& Modifier & & &
      \\
      Phenotype & alternative allele &  \multicolumn{3}{l}{Modification effect} & Reference\\
     \hline
     \multirow{2}{0.2\textwidth}{Coat color in Guinea pigs} & & \textit{pp} & \textit{P-} & & \multirow{2}{0.11\textwidth}{\cite{wright1927effects}} \\
     & \textit{C} → \textit{c}& Partial\tnote{1} & Dominant & \\
     \\

    \multirow{2}{0.2\textwidth}{DMI, hybrid male sterility in Drosophila} & & \textit{q\textsubscript{3}q\textsubscript{3}q\textsubscript{4}q\textsubscript{4}} &\textit{Q\textsubscript{3}q\textsubscript{3}q\textsubscript{4}q\textsubscript{4}} &  \textit{Q\textsubscript{3}q\textsubscript{3}Q\textsubscript{4}q\textsubscript{4}} & \multirow{2}{0.11\textwidth}{\cite{chang2010epistasis}}\\
     &\textit{Q\textsubscript{2}} → \textit{q\textsubscript{2}} & 
     Recessive & Partial\tnote{1} & Dominant & \\
     \\

    \multicolumn{3}{l}{\textit{Resistance-related alleles and their dominance modifiers}} \\
     \multirow{2}{0.2\textwidth}{Pathogen resistance in flies} & & \textit{M-} & \textit{mm} & & \multirow{2}{0.14\textwidth}{\cite{mckenzie1988diazinon}}\\
     &\textit{R} → \textit{S} & Neutral & Dominant & &\\
     \\
     \multirow{4}{0.2\textwidth}{Pathogen resistance in water fleas} & & \textit{- - E-} & \textit{ccee} & \textit{C-ee} & \multirow{4}{0.11\textwidth}{\cite{ameline2021two}}\\
     &\textit{B} → \textit{b} & Recessive &	Dominant &	Neutral & \\
     & & \textit{E-} & \textit{ee} \\
     & \textit{C} → \textit{c} &	Recessive & 	Dominant \\
     \\

    \multicolumn{3}{l}{\emph{Lineage-, sex-, or mating-specific dominance}} \\

    \multirow{2}{0.19\textwidth}{Spawning migration timing in salmon} & & Pop. 1\tnote{2} & Pop. 2\tnote{2} & & \multirow{2}{0.11\textwidth}{\cite{thompson2020complex}}  \\
     & RoSA variants & Early\tnote{3} & Late\tnote{3} \\
     \\ 

    \multirow{2}{0.19\textwidth}{Maturity age in salmon} & & Female & Male & & \multirow{2}{0.11\textwidth}{\cite{barson2015sex}}  \\
     &  \textit{E} → \textit{L} (\textit{VGLL3}) &	Later\tnote{3} & Earlier\tnote{3}  \\
     \\

     \multirow{2}{0.19\textwidth}{Reproductive success in seed beetles } & & Female & Male & & \multirow{2}{0.12\textwidth}{\cite{grieshop2018sex}}  \\
     & Autosome & Recessive (Dominant) & Dominant (Recessive) \\
     \\
     
     \multirow{2}{0.19\textwidth}{Migratory tendency in rainbow trout} & & Female & Male & & \multirow{2}{0.11\textwidth}{\cite{pearse2019sex}}  \\
     & \textit{A} → \textit{R} &	Dominant & Partial\tnote{1} \\
     \\

    \multirow{2}{0.19\textwidth}{Body size in seed beetles } & & Female & Male & & \multirow{2}{0.11\textwidth}{\cite{kaufmann2021rapid}}  \\
     & Autosome & High\tnote{1} & Low\tnote{1}  \\
     \\
     \multirow{2}{0.19\textwidth}{Fitness in yeast} & & BY\textit{\textbf{a}} / 3S\textit{$\alpha$}\tnote{4}& 3S\textit{\textbf{a}} / BY\textit{$\alpha$}\tnote{4}& & \multirow{2}{0.11\textwidth}{\cite{matsui2022interplay}}  \\
     & Three loci  &	Low\tnote{1} & Par.\tnote{1}~~or Over\tnote{1}  \\
     
    \hline

    \end{tabular}
    \begin{tablenotes}\footnotesize
        \item[1] Partial and Par. are partial dominance; Low is nearly recessive; High is nearly dominant; Over is over-dominance.
        \item[2] Pop. 1 is the population in the Klamath basin; Pop. 2 is the population in the Sacramento basin.
        \item[3] The heterozygote's trait is more similar to the trait of one homozygote than the other homozygote.
        \item[4] The full genotype of BY\textit{\textbf{a}} / 3S\textit{$\alpha$} is BY \textit{MAT\textbf{a}} / 3S \textit{MAT$\alpha$}; 3S\textit{\textbf{a}} / BY\textit{$\alpha$} is 3S \textit{MAT\textbf{a}} / BY \textit{MAT$\alpha$}. \textit{MAT\textbf{a}} and \textit{MAT$\alpha$} are two mating loci in yeast \cite{hanson2017evolutionary}.
    \end{tablenotes}
    \end{threeparttable}

    \label{varDomi}
\end{table}


\subsection{Variable dominance of inter-specific introgressions causing hybrid male sterility in Drosophila}
\label{introgressionSterility}

In an experimental study carried out in the \textit{Drosophila pseudoobscura} species group, the severity of hybrid incompatibility was found to depend on the dominance of the inter-specific introgressions. This study aimed to dissect the genetic basis of male hybrid sterility by examining introgressions derived from \textit{D. persimilis} at three previously identified quantitative trait loci (QTLs; \textbf{Q2}, \textbf{Q3}, \textbf{Q4}) in the genetic background of its sister species, \textit{D. pseudoobscura bogotana} (Figure \ref{domiDMI}A) \cite{chang2010epistasis}. Different levels of male hybrid sterility were observed for individual introgressions. Homozygous introgression of Q2 or Q3 variants from \textit{D. persimilis} resulted in complete sterility. However, when the \textbf{Q2} or \textbf{Q3} loci had a heterozygous \textit{D. persimilis} / \textit{D. p. bogotana} genotype, minimal sterility was observed ($3.8\%$ sterile males with single-copy Q2 from \textit{D. persimilis}; $0\%$ with single-copy Q3 from \textit{D. persimilis}). This indicates that individual Q2 and Q3 alleles of \textit{D. persimilis} origin are recessive in the genetic background of \textit{D. p. bogotana}. Neither heterozygous nor homozygous Q4 introgression induced any sterility, indicating that individual Q4 introgressions of \textit{D. persimilis} origin have no effect on hybrid male sterility in \textit{D. p. bogotana}. 

When two or more of the recessive-acting or no-effect introgressions from \textit{D. persimilis} at these QTLs were combined in the genome of \textit{D. p. bogotana}, the proportion of sterile males significantly increased to as high as $94.3\%$ \cite{chang2010epistasis}. This indicates that the con-specific alleles from \textit{D. persimilis} modify each other to be less recessive, allowing for at least one of them to interact negatively with alleles in \textit{D. p. bogotana}. When assigning \textbf{Q2} as the focal locus, sterility of single-copy introgression Q2 increased with the number of additional introgressions, reflecting increasing dominance observed at \textbf{Q2} (Table \ref{varDomi}) \cite{chang2010epistasis}: Q2 in a pure \textit{D. p. bogotana} background is recessive; Q2 is partially dominant in the presence of single-copy Q3; and Q2 is completely dominant in the presence of single-copy Q3Q4. Notably, single-copy Q3 and single-copy Q3Q4 introgressions have minimal effects on sterility (Figure \ref{domiDMI}A). Thus, dominance of Q2 was enhanced by the con-specific introgressions Q3 and Q4. 

Previous models of introgression and hybrid incompatibilities have usually assumed constant dominance. The work described above shows that presumed recessive-acting introgressed alleles became dominant for male hybrid sterility in the presence of con-specific alleles. This variable dominance should lead to strong negative effects in early-generation hybrids (thus, helping maintain species barriers) that could rapidly be ameliorated through backcrossing and subsequently allow individual introgressions to spread. New models are therefore needed to quantify the effect of variable dominance on the fate of hybrids and expected patterns of introgression in the short and long term.   

\subsection{Variable dominance at resistance-related loci}

As a specific type of dominance modification, the presence of alleles at other loci can suppress the observed effect of a dominant allele at a focal locus, rendering the focal allele neutral and thereby erasing dominance. This phenomenon is exemplified at loci associated with pathogen or insecticide resistance (Table \ref{varDomi}). Resistance evolution is often associated with a fitness cost for the organism in the absence of the selection pressure driving resistance. As resistance evolves, compensatory mutations at other loci emerge or rise to high frequency that ameliorate this fitness cost (reviewed in \cite{bass2017does, baucom2019evolutionary}). One example in which compensatory evolution modified dominance is the case of resistance to the insecticide diazinon in the fly \textit{Lucilia cuprina}. Resistance occurred through a dominant mutation at locus \textbf{R} \cite{mckenzie1988diazinon}. Initially, carriers of at least one resistance allele also suffered from a developmental abnormality that increased the asymmetry of the left and right sides of the organism. However, with the emergence of a compensatory mutation at another locus \textbf{M}, the organism could largely correct this asymmetry and mitigate the negative effects associated with the resistance allele both in the presence and the absence of the insecticide. Thus, the resistance allele, which is dominant for the trait "resistance", appeared initially dominant for the trait "asymmetry". However, its dominance (and its effect altogether) with regard to "asymmetry" was neutralized by the appearance of the compensatory allele at locus \textbf{M}. Since the compensatory allele increases the fitness of resistance allele carriers both in the presence and the absence of the insecticide, dominance at locus \textbf{R} will appear variable (albeit at different degrees) in both environments.

The complexity of dominance is also visible in the genetics underlying resistance to the pathogen \textit{Pasteuria ramosa} in the planktonic crustacean \textit{Daphnia magna} \cite{bento2017genetic,ameline2021two}. The authors studied resistance against three pathogen strains, C1, C19 and P20. Genome-wide association studies revealed that resistance is controlled by at least three loci, \textbf{B}, \textbf{C}, and \textbf{E}. Alleles at the \textbf{B} and \textbf{C} loci can facilitate resistance, where individuals with genotype $bbcc$ are susceptible to all three pathogen strains. Alleles $B$ and $C$ appear dominant, but their expression is contingent upon the presence of a modifying allele at locus \textbf{E}. In the $ee$ background, genotypes with a $B$ allele ($B-$) show resistance against C19 and P20, and genotypes with a $C$ allele ($C-$) show resistance against all three pathogen strains. However, when the genotype at locus \textbf{E} is $EE$ or $Ee$, the resistance effects of alleles $B$ and $C$ against P20 are suppressed. However, allele $E$ does not suppress the resistance effect of $B$ and $C$ alleles against C19. Moreover, in the absence of the resistance allele $C$ and the masking allele $E$ (i.e., in the background of genotype $ccee$), the $B$ allele confers resistance to C19 and P20. Here, the $C$ allele modulates dominance at the \textbf{B} locus, which further complicates the picture of dominance in this system. This example shows that both dominance and its modification can be phenotype-specific; in this example, the phenotype is resistance to a specific pathogen strain.

\subsection{Potential dominance modifiers in lineage-, sex- or mating-specific dominance}\label{fish}

Detecting dominance modifiers is challenging for at least two reasons. Firstly, at the population level, the modification effect may be too weak to detect because the interaction induces subtle changes in allele or genotype frequencies \cite{wright1934physiological}. Secondly, the higher-order effect can be misinferred as an additive effect \cite{crow2009introduction, huang2016genetic, li2023rapid} (Box \hyperlink{polygenic}{2}). The existence of dominance modifiers is likely when the observed dominance at a focal locus varies across different lineages, sexes, or mating types (Table \ref{varDomi}). For example, lineage-specific dominance was reported from two ecotypes of Chinook salmon with different spawning migration timing \cite{thompson2020complex}. In this case, a genomic region called RoSA, spanning approximately $30$-kb, is nearly perfectly associated with migration timing. Heterozygotes at RoSA in lineages found in the Klamath basin resemble the early-migrating ecotype, whereas heterozygotes in lineages found in the Sacramento basin resemble the late-migrating ecotype. Notably, without further experimentation, it is impossible to conclude whether the observed dominance reversal is lineage- or environment-specific. If dominance is indeed lineage-specific in this system, the interaction between the two alleles at RoSA is likely influenced by lineage-specific modifiers at other loci. 

Dominance may also differ between the two sexes. Such sex-specific dominance is believed to alleviate sexual conflict by resolving conflicting fitness interests between sexes. Sex-specific dominance allows for a favored allele to be expressed and an unfavored allele to be silenced in each sex, respectively \cite{spencer2016evolution}. Empirical evidence supporting sex-specific dominance comes from studies of the genetic loci underlying life-history traits, which often serve as indicators of fitness (Table \ref{varDomi}). Examples include the age of maturity in Atlantic salmon \cite{barson2015sex}, reproductive success in seed beetles \cite{grieshop2018sex}, migration rate in rainbow trout \cite{pearse2019sex}, and body size in seed beetles \cite{kaufmann2021rapid}. These studies have demonstrated that the dominance of certain genomic fragments involved in these traits varies between males and females. Here, sex-specific dominance modifiers might not be located on the sex chromosomes or near sex-determining loci. However, their necessary association with sex determination may make sex-specific dominance modifiers easier to detect than other dominance modifiers.

It is of great interest to estimate the dominance coefficient of naturally-occurring or mutagen-induced mutations \cite{mukai1972mutation, orr1991test, agrawal2011inferences, manna2011fitness, yang2017incomplete}. In approaches to infer dominance, dominance modifiers of newly arising mutations are often overlooked or not discernible, especially when new mutations are not specifically linked to a particular trait and exhibit weak associations with the trait. ``Evolve and resequence" (E\&R) experiments have shown some power to estimate dominance coefficients and to identify modifiers, \textit{e.g.}, in yeast \cite{matsui2022interplay} (Table \ref{varDomi}) . In E\&R experiments, experimental populations are subjected to competition and selection over multiple generations. By analyzing changes in allele frequencies during the competition, one can infer the fitness of different genotypes and estimate the dominance coefficients of polymorphic loci (e.g., SNPs) associated with fitness. For instance, E\&R experiments were carried out in populations generated by crossing two haploid yeast strains (3S and BY) of two mating types (\textit{MAT\textbf{a}} and \textit{MAT$\alpha$}), creating a diverse set of recombinants,  and subsequently letting the resulting diploid yeast populations evolve \cite{matsui2022interplay}. The authors identified multiple modifiers that influenced the dominance at focal loci, with the mating type alleles being the most important modifier. In the BY \textit{MAT\textbf{a}} / 3S \textit{MAT$\alpha$} genetic background, there was little dominance at three focal loci (i.e., alleles appeared additive). However, when the mating type was 3S \textit{MAT\textbf{a}} / BY \textit{MAT$\alpha$}, dominance at focal loci became stronger, and occasionally, over-dominance was observed. Thus, through the dependence of dominance on the mating locus, the yeast study showed similarities to the above-discussed cases of sex-specific dominance. 

\subsection{Challenges in understanding dominance and multi-locus interaction} \label{chanllengeVar}

The examples above (summarized in Table \ref{varDomi}) demonstrate the tight connection between dominance and epistasis. Two key insights warrant attention. First, there exists a huge gap between observing dominance and our understanding of its genetic basis, a situation reminiscent of the study of epistasis \cite{domingo2019causes}. It is important to understand better whether (and how frequently) observed dominance results from dominance modifiers through epistatic interactions or merely from a non-additive relationship between the genotype at a focal locus and the phenotype or fitness. To untangle these effects empirically, precise control of the genetic background is necessary, which can be challenging \cite{lachance2013genetic, chandler2017well}. 

Second, only strong non-additive effects can be identified with current approaches. However, the majority of dominance modification effects might be weak. It is possible that epistatic modification of dominance is a widespread phenomenon that affects many loci in the genome. The case studies presented here may represent only the tip of the iceberg, which stood out because of their strong effects on important phenotypes. Here, modeling and theory may help identify the evolutionary consequences and signatures of widespread modification effects.

\FloatBarrier
\section{Modelling dominance}

Although empirical work has shown as early as 1910 that the dominance of focal alleles can depend on their genetic background  \cite{bateson1910reports}, the variable nature of dominance tends to be neglected. Here, modeling could provide guidance on how to reveal and parameterize dominance. Theoretical models that incorporate variable dominance could be used to quantify its consequences on evolutionary dynamics and to predict when variable dominance should be considered in the study of adaptation and speciation.

\subsection{Dominance in hybrid incompatibility} \label{defineDMI}

\begin{figure}
    \centering
    \includegraphics[width=0.92\textwidth]{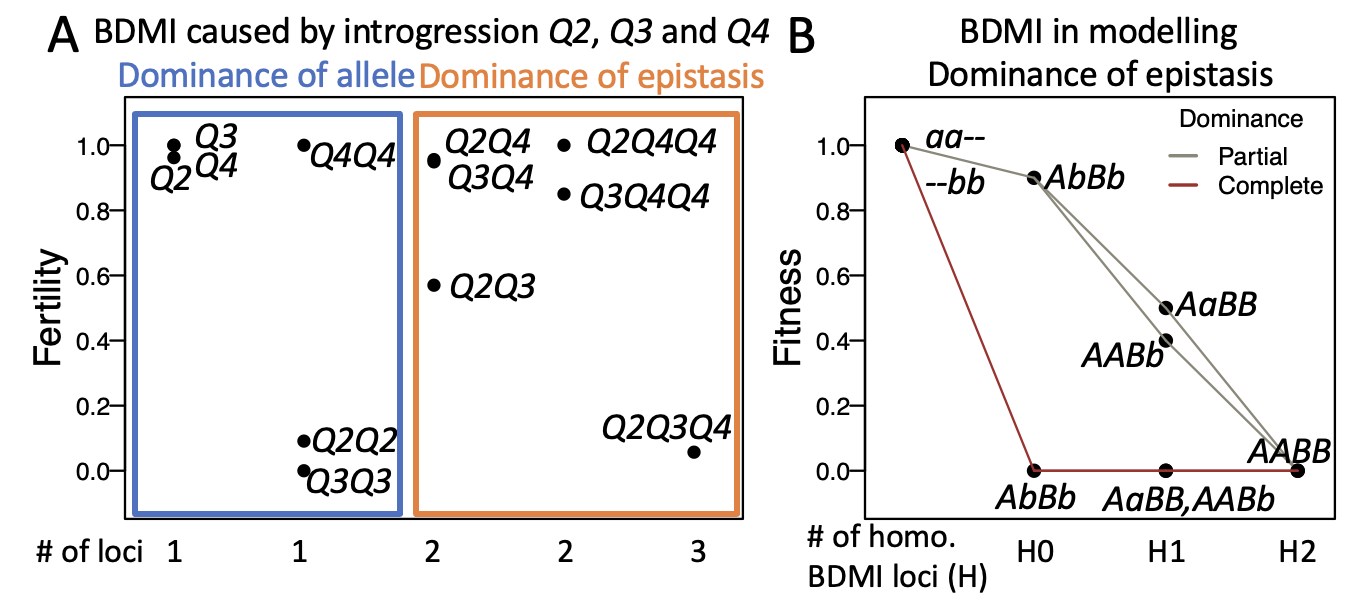}
    \caption{Dominance in genetic incompatibility. A. Dominance of inter-specific introgressions in the context of a BDMI. The fitness of three introgressions, measured by the fertility, was examined in the genetic background of \textit{D. pseudoobscura bogotana}. The introgressions were derived from its sister species \textit{D. persimilis} at three QTL (\textbf{Q2}, \textbf{Q3}, \textbf{Q4}) (see also \ref{introgressionSterility}) \cite{chang2010epistasis}. Fertility varies depending on the number of introgressions, demonstrating variable dominance of introgressed alleles (see Section \ref{introgressionSterility}). The dominance of individual alleles can be estimated independently for single introgressions (in blue) or by considering the dominance of epistasis (in orange). B. Fitness landscapes of a two-locus Bateson-Dobzhansky-Muller incompatibility (BDMI). Here, hybrid incompatibility occurs between alleles $A$ and $B$, leading to reduced fitness. The fitness landscape illustrates the fitness values associated with different genotype combinations at the two loci. The numbers of homozygous incompatible loci range from zero to two (H0, H1 and H2), for which dominance of the interaction must be specified. Note that this diagram assumes no dominance at each individual locus.}
    \label{domiDMI}
\end{figure}

The quantification of dominance becomes increasingly complicated when considering its role in multi-locus interactions, because not only the dominance at each individual locus has to be specified but also the dominance of the interaction. This complication can be exemplified in the two-locus Bateson-Dobzhansky-Muller incompatibility (BDMI) model \cite{maheshwari2011genetics}, which is the classic model to explain the evolution of hybrid incompatibility in allopatric populations. The BDMI model features the incidental evolution of strong negative epistasis between two (or more) loci that can be individually neutral, eventually resulting in hybrid sterility or inviability (indicating reduced fitness) when two diverged populations or species hybridize. 

Two-locus BDMIs can be classified into three categories based on the number of homozygous sites necessary to affect hybrid fitness significantly, ranging from zero to two (H0, H1 and H2 genotypes; Figure \ref{domiDMI}B) \cite{turelli2000dominance}. Whether and how strongly H0 and H1 genotypes are suffering from reduced fitness determines the dominance of the BDMI. Theoretical work has shown that especially whether first-generation (i.e., double heterozygote, H0 genotypes) hybrids are susceptible to the incompatibility -- i.e., whether the BDMI is recessive (no reduced fitnes of H0 genotypes) or (co-)dominant (H0 genotypes suffer from reduced fitness) -- greatly affects evolutionary dynamics and the probability of a hybrid swarm to survive \textit{e.g.}, \cite{bank2012limits, blanckaert2018search, blanckaert2023search}. 

In first-generation hybrids carrying a BDMI, heterogametic offspring (chromosomes XY or ZW) tend to suffer more from a BDMI than homogametic offspring (XX or ZZ), a phenomenon known as Haldane's rule \cite{haldane1922sex}. Haldane's rule can be partially explained from a dominance perspective \cite{turelli1995dominance, schilthuizen2011haldane}. In homogametic hybrids, if a BDMI involves two recessive incompatible alleles, the incompatibility is masked. In contrast, a recessive incompatible allele is always expressed in hemizygous individuals. Consequently, BDMI between hemizygous and heterozygous loci are expected to be more severe than between two heterozygous loci. Haldane's rule implies that the dominance of the individual incompatible alleles is related to the dominance of the incompatibility \cite{turelli1995dominance}, and that BDMIs tend to be recessive. 

In summary, the definition of dominance in the context of multi-locus interaction involves considering dominance at each locus and the dominance of epistasis, \textit{e.g.}, \cite{hvala2018signatures, dagilis2019evolution}, which results in a large number of parameters to be defined in a model. Moreover, as the empirical examples in Section \ref{domiEpi} showed, assuming constant dominance coefficients may not adequately capture the complexity of dominance in the context of the larger genetic background of diverse populations. Indeed, epistatic interactions with modifier alleles may render some alleles dominant in some genetic backgrounds and recessive in others (Figure \ref{domiDMI}A). Determining minimal and generalizable models that capture this complexity and its consequences is challenging. 

\subsection{Integrating constant and variable dominance in a three-locus epistatic model} \label{domiModel}

\begin{figure}[htp]
    \centering
    \includegraphics[width=0.9\textwidth]{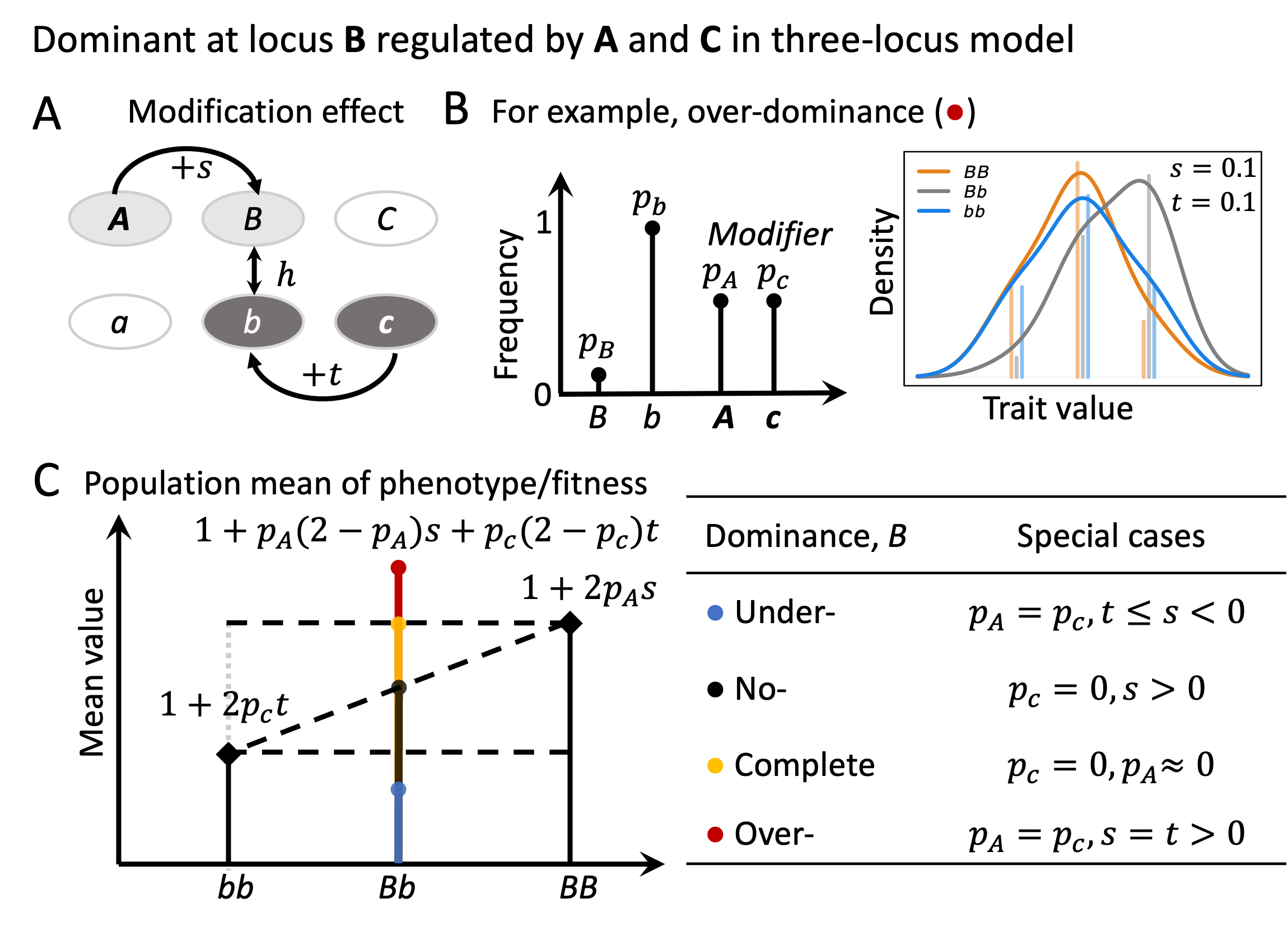}
    \caption{Variable dominance resulting from multi-locus interaction. A. A three-locus model illustrates how variable dominance is caused by epistasis \cite{wade2001epistasis}. Dominance at locus \textbf{B} is influenced by the allelic state at its modifier loci, \textbf{A} and \textbf{C}. Under the assumption of no position effects and dominance of the modifying alleles, the presence of one-copy allele $A$ changes the effect of one-copy allele $B$ by $s$, and the presence of allele $c$ changes the effect of allele $b$ by $t$. Depending on the allele frequencies at each of the three loci, the entire spectrum of dominance can be observed. B. A scenario in which over-dominance is observed under the model from A. When the alleles at the focal locus (\textbf{B}) and its modifiers (\textbf{A} \& \textbf{C} ) segregate in a population, the effect of each genotype at locus \textbf{B} is often estimated from the trait distribution. The mean effect of each genotype at locus \textbf{B} can be computed as a function of the modification effect ($s,t$) and the frequency of modifier alleles ($p_A, p_c$). In this toy model, only three trait values can be observed in the population. The trait distribution for each genotype ($BB$, $Bb$, $bb$) is smoothed. C. Four frequency-dependent dominance scenarios at locus \textbf{B}, computed by comparing the mean effect of genotype $Bb$ and $BB$, indicated by four colors in the lower left panel.}
    \label{wadeThree}
\end{figure}


From a modeling perspective, a constant dominance coefficient is captured by the interaction of two alleles at a focal locus. The position effect indicates the existence of a cis-regulatory dominance modifier allele that influences the dominance at the focal locus, creating a two-locus epistatic scenario (Figure \ref{domiClassical}C). Here, the observed dominance at the focal locus varies depending on the allelic state at another locus. If each allele at the focal locus has its own modifier alleles at different loci, the regulation of dominance becomes more complex. Wade \cite{wade2001epistasis} proposed a model that captures this scenario, which is illustrated in Figure \ref{wadeThree}, where two loci \textbf{A} and \textbf{C} regulate a focal locus \textbf{B}. At locus \textbf{B}, the effect of allele $A$ changes the effect of allele $B$ to $1+s$, and the effect of allele $c$ changes the effect of allele $b$ to $1+t$. In a polymorphic population, as the allele frequencies at locus \textbf{A} and \textbf{C} change, the observed dominance at the focal locus \textbf{B} also changes. When the modifier loci are monomorphic, two special cases arise. When both $A$ and $c$ are fixed in the population, the alleles at locus \textbf{B} are perceived to be additive. Conversely, when one of the modifying alleles is fixed in the population and the other lost, the modified allele at locus \textbf{B} is observed to be dominant. 

Thus, under Wade's model, observed dominance coefficients, which can include incomplete, under-, and over-dominance, vary with the allele frequencies at their modifier loci \cite{omholt2000gene, wade2001epistasis}. The dominance of heterozygotes $Bb$ at the focal locus is determined by the frequencies of interacting alleles ($A, c$) and their regulatory effects ($s, t$): the phenotype of $Bb$ is $1+p_{A}(2-p_A)s+p_{c}(2-p_c)t$, and the corresponding homozygous phenotypes are $1+2sp_{A}$ for $BB$ and $1+2tp_{c}$ for $bb$ (Figure \ref{wadeThree}). Here, over-dominance occurs when $p_A=p_c, s=t>0$; under-dominance occurs when $p_A=p_c, t\leq s<0$ (Figure \ref{wadeThree}). In the regime of over-dominance, heterozygotes are favored, thereby maintaining genetic polymorphism within a population. In the regime of under-dominance, disruptive selection will favor homozygotes and drive genetic differentiation between populations. Therefore, the variable dominance resulting from multi-locus interaction has important evolutionary consequences, which may shape genetic diversity and structure of populations.

The model proposed by Wade can be considered a minimal model of dominance through epistatic interaction. Already this minimal model contains three loci and two parameters without considering direct selection at any of the three loci. Moreover, we expect that the recombination rate between the interacting loci is an additional important parameter that affects evolutionary dynamics under this model. This highlights once more the complexity caused by dominance and multi-locus interaction, which poses challenges to both empirical and theoretical research.    

\subsection{Revisiting hybrid incompatibility and introgression in the light of Wade's model}\label{DMIdom}

We can now consider the fitness landscapes of hybrid incompatibility (and its dominance) in the context of variable dominance as modeled by Wade (Figure \ref{wadeThree}) \cite{wade2001epistasis,wade2002gene}. A two-locus BDMI involves at least two loci that interact. To incorporate Wade's model, each of these loci needs two accompanying modifier loci. For instance, assume a BDMI between allele \textit{B\textsubscript{1}} and \textit{B\textsubscript{2}} at loci \textbf{B\textsubscript{1}} and \textbf{B\textsubscript{2}}. The strength of negative epistasis, when a heterozygous genotype is included, is then determined by the dominance of the alleles \textit{B\textsubscript{1}} and \textit{B\textsubscript{2}}.  Modifier alleles for each allele at the focal loci determine the observed dominance: locus \textbf{A\textsubscript{1}} and \textbf{C\textsubscript{1}} modify the effect of  \textbf{B\textsubscript{1}}; \textbf{A\textsubscript{2}} and \textbf{C\textsubscript{2}} modify the effect of \textbf{B\textsubscript{2}}. When the dominance-regulating alleles (\textit{A\textsubscript{1}} and \textit{A\textsubscript{2}}) segregate at sufficiently high frequencies, the hybrid incompatibility will appear dominant (i.e., show effects on double heterozygotes). High frequencies of modifier alleles (as expected when these are near fixation in both parental populations) would thus increase the observed effect of hybrid incompatibilities during genetic mapping, potentially to the point that mapping becomes impossible because of the low fitness of first-generation hybrids. This presents a challenge to the identification of hybrid incompatibilities. Moreover, if a hybrid population persists, we hypothesize that there would be indirect selection against the modifier alleles (because they essentially make hybrids unfit), such that the observed dominance of the BDMI may decrease over time. Both of these patterns could contribute to the empirical finding that most identified BDMI loci appear to be recessive \cite{orr2000speciation,presgraves2021hybrid}. Further theoretical work is necessary to confirm these predictions. 

Interestingly, Wade's three-locus model captures the introgression effects on male sterility in \textit{Drosophila pseudoobscura} species group \cite{chang2010epistasis}. As shown in Figure \ref{domiDMI}A, it appears that the dominance of one introgression is amplified by the other conspecific introgression, leading to male hybrid sterility. Seen in the light of Wade's model, one could assume that Q3 modifies Q2. Individually, introgressions Q2 and Q3 exhibit recessive effects on sterility. However, when the two introgressions are present in the same genome, they synergistically enhance sterility. This was hypothesized to be caused by the negative interaction between one introgression (Q2) and at least one unidentified allele at other loci of the host genome \cite{chang2010epistasis}. In addition, the partial effects from dominance modifiers also indicate that more than two loci epistatically regulate the dominance at locus \textbf{Q2}. In summary, Wade's model, extended to BDMIs, reveals how the interplay between dominance and epistasis may alter the dominance of introgressions and hybrid incompatibilities and thereby influence inter-species gene flow.

\FloatBarrier
\subsection{Dominance in fitness landscapes}
\label{fl}
\begin{table}[htp]
    \centering
    \caption{Requirements for fitness landscape models capturing different dominance scenarios}
    \begin{tabular}{p{0.22\textwidth} p{0.22\textwidth} p{0.22\textwidth} p{0.22\textwidth}}
    \hline
    Dominance & Recombination & Modifier alleles & Number of loci ($L$) \\
    \hline
    One locus &  NA & NA  & 1 \\
    Position effect & $r\approx0$ & 2 & 2 \\
    Wade's model & $[0,0.5]$ & $\geq 2$ & $\geq 3$ \\
    Two-locus DMI & $[0,0.5]$ & $\geq 4$ & $\geq 6$ \\
    Polygenic traits & $[0,0.5]$ & Multiple & Multiple \\
     \hline
    \end{tabular}
    \label{tableWade}
\end{table}

To highlight what is necessary to incorporate dominance and gene-gene interactions into fitness landscape models, we compiled the minimum number of loci required to exhaustively parameterize each dominance scenario following Wade's model (Table \ref{tableWade}). These scenarios include constant one-locus dominance determined by the differences of additive effects between alleles at a focal locus, and variable dominance arising from gene-gene interactions occurring either in \textit{cis} or \textit{trans}. Moreover, observed dominance levels and the resulting evolutionary dynamics will be influenced by various factors such as the recombination rates between the focal locus and its modification loci, frequencies of modifier alleles, and their modifying effects. From a fitness landscape perspective, we might ask: which relative contributions of additive effects and epistasis in a fitness landscape resemble the distribution of dominance and epistasis that are observed in polymorphic populations or between species?

Empirical distributions of inferred dominance seem to be similar across many species \cite{mukai1972mutation, orr1991test, agrawal2011inferences, manna2011fitness, yang2017incomplete}. Previous research used Fisher's Geometric model, which is a model that maps phenotype to fitness, to explain the dominance of mutations, successfully reproducing empirical distributions \cite{manna2011fitness}. However, to understand the genetic basis, it is worth integrating dominance into genetically explicit fitness landscape models in the future. By extending fitness landscape research to diploid populations subject to the consequences of dominance and epistasis, we might gain a broader understanding of the role of dominance in shaping evolutionary outcomes. For example, we recently studied the rapid adaptation of a recombining haploid population driven by standing genetic variation after a sudden environmental change \cite{li2023rapid, Amado2023STUN}. We showed that epistasis leaves a signature of positive frequency-dependent selection \cite{li2023rapid} and that alleles could be maintained at low frequencies for long times \cite{Amado2023STUN}. We expect that dominance may add further complexity to the evolutionary dynamics, for example, by extending segregation times of negatively interacting alleles, as argued in section \ref{DMIdom}. Moreover, models of fitness landscapes that allow for tunable dominance in addition to epistasis may ultimately serve as alternative null models in studies of complex polygenic traits, which usually assume additivity.

\section{Roads and obstacles to quantifying dominance and multi-locus interactions}

Inference of dominance in natural populations is challenging due to its statistically low heritability and a large number of potential dominance scenarios \cite{palmer2023analysis}. Multi-locus interactions are similarly difficult to infer \cite{hansen2006evolution}. However, pinpointing both is a necessary part of reliably identifying the genetic basis of many traits. Also in examples from medical genetics, dominance at focal disease-causing was demonstrated to vary due to the impact of genetic interactions, ranging from complete recessivity to complete dominance\cite{deciphering2017prevalence, veitia2018mechanisms, kingdom2022incomplete}. Here, deciphering the role of dominance and multi-locus interactions is essential for reliable genetic diagnosis \cite{zschocke2023mendelian}. Developing statistical methodologies and constructing new null models to capture the dominance and epistasis remains a formidable challenge (Box \hyperlink{polygenic}{2}; section \ref{fl}). Despite the challenges, development of simulation-based methods, aided by model predictions and incorporation of knowledge from systems and molecular biology, is a potential avenue for further investigation.

In this review, we have argued that dominance may vary with the genetic background. Dominance may also vary across environments, as was briefly mentioned in Section \ref{fish}. A notable example is seen in the \textbf{Ace} locus, which regulates insecticide resistance in mosquitoes \cite{bourguet1996dominance}. The dominance of the resistance allele at \textbf{Ace} varies from almost complete dominance to near-recessivity across different environments. Regarding model development, variable environments may also be captured by fitness landscape (or seascape) models \cite{mustonen2010fitness, bank2022epistasis}, while again creating further complexity. Quantifying the causes and consequences of variation of non-additive effects with the environment deserves further research to identify its role in polygenic trait evolution.

\section{Concluding Remarks}

Variable dominance, which can arise from multi-locus or genotype-environment interactions, affects the fate of populations in the short and long term. During evolution, variable dominance may promote or hinder maintenance of genetic diversity, or 
drive population differentiation. Moreover, dominance of the multi-locus interactions themselves affects the spread and fixation of alleles and the fate of hybrid populations \cite{bank2012limits,lachance2011population, blanckaert2018search, blanckaert2023search}. As of today, many questions remain regarding the prevalence and genetic and basis of dominance, and the evolutionary consequences of dominance and multi-locus interactions  (see Outstanding Questions). Addressing these questions requires the development and integration of new models, statistical approaches, and genetic and molecular studies. 

\FloatBarrier
\section*{Acknowledgements}
The authors thank Chau-Ti Ting for advice on how to structure this manuscript, Alexandre Blanckaert for suggesting literature, the THEE lab members for contributing feedback and ideas, and Vitor Sousa for discussions of epistasis and dominance. We thank Meike Wittmann and Vitor Sudbrack for their helpful comments and suggestions after reading the manuscript. We thank Maria Smit and three anonymous reviewers for their constructive reviews and thoughtful comments during the review process. This work was supported by funding from ERC Starting Grant 804569 (FIT2GO) and HFSP Young Investigator Grant RGY0081/2020 to CB.  

\section*{Declaration of Interests}

The authors declare no competing interests.

\section*{Declaration of Generative AI and AI-assisted technologies in the writing process}

During the preparation of this work, the authors have used ChatGPT in order to ensure adherence to English language syntax conventions. After using this tool/service, the authors reviewed and edited the content as needed. The authors take full responsibility for the content of the publication.

\bibliography{EpiDomiRefs} 

\end{document}